\documentclass[12pt]{article}
\usepackage{graphicx}
\usepackage{amsmath}
\usepackage{tikz}
\usepackage{amsthm}
\usepackage{subcaption}
\usepackage{amsfonts}
\usepackage{float}   
\date{}
\usepackage{xcolor}
\usepackage{tabularx}
\usepackage{booktabs} 
\usepackage{multirow}
\usepackage{dsfont}
\usepackage{hyperref}
\usepackage{comment}
\DeclareMathOperator*{\argmax}{arg\,max} 
\DeclareMathOperator*{\argmin}{arg\,min} 
\usepackage{enumerate}
\usepackage{natbib} %comment out if you do not have the package
\usepackage{url} % not crucial - just used below for the URL 
\newcommand{\blind}{0}

\begin{document}
\def\spacingset#1{\renewcommand{\baselinestretch}%
{#1}\small\normalsize} \spacingset{1}

%%%%%%%%%%%%%%%%%%%%%%%%%%%%%%%%%%%%%%%%%%%%%%%%%%%%%%%%%%%%%%%%%%%%%%%%%%%%%%

\if0\blind 
{
  \title{\bf Benchmark of Likelihood-Free Inference Methods based on Neural and Optimal Transport Approaches}
  \author{Samira Aka\hspace{.2cm}\\
    LSCE, 
 Universit\'e Paris Saclay \& ESSEC Business School CREAR\\
    and \\
    Marie Kratz \\
    ESSEC Business School, IDO Dep. and CREAR\\
    and \\
    Philippe Naveau 
    \\ Laboratoire\,des\,Sciences\,du\,Climat\,et\,de\,l’Environnement, CNRS-CEA-UVSQ-IPSL}
  \maketitle
} \fi

\if1\blind
{
  \bigskip
  \bigskip
  \bigskip
  \begin{center}
    {\LARGE\bf Title}
\end{center}
  \medskip
} \fi

\bigskip
\begin{abstract}
Simulation-based inference (SBI) has become an increasingly important framework for parameter estimation in models for which simulation is feasible, including cases where likelihood evaluation is unavailable or costly. While recent work has introduced benchmark frameworks to compare likelihood-free methods, these studies often do not account for structural features such as heavy-tails or discreteness.

In this article, we investigate how the performance of likelihood-free inference methods depends on these structural properties. We consider four approaches: MLE, NBE, EOT and AW--NBE and evaluate them using simulations.
This study highlights the importance of carefully selecting evaluation tools in the presence of extremes and discrete data.
\end{abstract}

\noindent%
\vfill
\newpage
%\tableofcontents
\newpage
%\spacingset{1.8} % DON'T change the spacing!

\section{Introduction}
\label{sec:intro}

Simulation-based inference (SBI) has become a major framework for parameter inference in models for which the likelihood is intractable, but from which simulation remains feasible (\cite{Sisson2018,Cranmer2020,Frazier2020,BrehmerCranmer2021}). Early likelihood-free approaches, most notably Approximate Bayesian Computation (ABC), approximate the posterior distribution by retaining parameter values that generate simulated data sufficiently close to the observation, typically through summary statistics and tolerance thresholds (\cite{Beaumont2002,Beaumont2009,Beaumont2010}). Although widely used, these methods are known to be sensitive to the choice of summary statistics, which directly impacts the quality of the resulting approximation (\cite{Beaumont2002,Burr2013ABCsummaries}). To alleviate some of these limitations, more recent developments have shifted SBI toward neural approaches, in which a neural network is trained offline on simulated pairs (parameters, observations) and subsequently enables fast inference for new observations. Depending on the target quantity, one distinguishes Neural Posterior Estimation (NPE), which directly approximates the posterior distribution (see e.g. \cite{Papamakarios2016NPE,Lueckmann2017SNPE}), Neural Likelihood Estimation (NLE, or SNLE in its sequential form), which learns a surrogate likelihood and combines it with Bayes' formula (\cite{Drovandi2020ABC,Papamakarios2019}), and Neural Ratio Estimation (NRE), which estimates the likelihood-to-evidence ratio through a classification-based formulation (\cite{Durkan2019,Hermans2020}). These neural-based approaches now constitute a central class of methods in modern SBI, and are typically evaluated through their ability to approximate the posterior distribution, as well as through calibration and predictive performance criteria \cite[see, e.g.][]{Huang2023RobustSBI,SpurioMancini2023SBIModelComparison}. Benchmarking studies such as \cite{Lueckmann2021} provide systematic comparisons of SBI algorithms across a collection of standardized simulator tasks, by assessing how well the resulting approximate posteriors match reference posteriors using statistical discrepancy measures such as classifier two-sample tests and maximum mean discrepancy. In a complementary direction, \cite{deistler2025simulationbasedinferencepracticalguide} offer a practical perspective on the full SBI pipeline, covering simulation design, diagnostics, calibration assessment, and posterior validation.\\
In contrast to these works, which primarily focus on posterior approximation, we consider here the problem of point estimation. In fact, these two perspectives correspond to distinct inferential objectives. Standard SBI methods target the full conditional distribution of parameters knowing data, whereas point estimation methods aim directly at a single decision-oriented quantity, typically an estimator of parameters minimizing a predefined loss. Classical approaches to point estimation are often formulated as extremum estimators, including maximum likelihood estimation (MLE) and minimum distance-based estimation \cite[see, e.g.][]{NeweyMcFadden1994}. More recently, \cite{sainsbury2024neural} formalized the framework of Neural Bayes Estimators (NBE), which are trained to approximate Bayes estimators rather than posterior distributions. This distinction is essential for benchmarking. In contrast with standard SBI benchmarks, which compare posterior distributions, we focus here on methods that return point estimates. This yields a coherent comparison framework and avoids introducing an additional layer of variability through posterior distribution summaries. Our benchmark therefore compares four estimation strategies: MLE, taken as a reference under correct specification, NBE for  point estimation; entropic optimal transport estimation (EOT), which replaces likelihood criteria by a regularized transport discrepancy between distribution (here empirical) \cite[see, e.g.][]{cuturi2013sinkhorn,Genevay2018}, and the adaptive Wasserstein neural Bayes estimator (AW--NBE) \cite[see, e.g.][]{Aka2026}, which combines an NBE initialization with an optimal transport refinement.

Our main contributions are threefold. First, we provide a unified benchmark that brings together likelihood-based and likelihood-free methods, which are typically studied separately, and offer a systematic comparison across these distinct inference paradigms. Second, we shift the focus from posterior approximation to point estimation, providing a coherent framework to compare methods targeting different inferential objectives. Third, we investigate how structural properties of the data, including heavy tails, discreteness, and model misspecification, affect the performance of the chosen inference methods.

The purpose of this article is therefore not to introduce another inference algorithm, but to examine systematically how different estimation paradigms behave across controlled statistical settings. We compare MLE, NBE, EOT, and AW--NBE under light-tailed, heavy-tailed, and discrete settings, both in correctly specified and misspecified scenarios, and evaluate them through numerical error criteria.

The remainder of this article is organized as follows: Section~\ref{sec:framework} describes the benchmark framework, including the methods under comparison, the data-generating 
processes, and the evaluation metrics. From Section~\ref{sec:gauss} to Section~\ref{sec:doublePareto}, we present the results obtained by the different inference methods for each model under study. Section~\ref{sec:conclusion_benchmark} provides an Overall: discussion and key takeaways on the relative strengths and limitations of the considered estimation strategies.
Throughout the benchmark, $\boldsymbol{\theta}$ denotes the parameter vector of the model
currently under consideration.\\[1ex]
\noindent\textbf{Notation.}
We denote by $P_{\boldsymbol{\theta}}$ the distribution of the observations under
parameter $\boldsymbol{\theta} \in \Theta$, and by $\pi(\boldsymbol{\theta})$ the prior
distribution used for neural training.
Given an observed sample $\boldsymbol{x}_1, \ldots, \boldsymbol{x}_m$, we write
$\widehat{P}_m = \frac{1}{m}\sum_{i=1}^m \delta_{\boldsymbol{x}_i}$
for the corresponding empirical measure.
For each $\boldsymbol{\theta} \in \Theta$, we simulate a sample $(y_i)$ of size $m_s$
from $P_{\boldsymbol{\theta}}$ and denote its empirical measure by
$\widehat{P}_{m_s,\boldsymbol{\theta}} = \frac{1}{m_s}\sum_{j=1}^{m_s}
\delta_{\boldsymbol{y}_j^{(\boldsymbol{\theta})}}$.
Estimators of $\boldsymbol{\theta}$ are denoted
$\widehat{\boldsymbol{\theta}}^{\mathrm{Method}}$, where the superscript identifies
the estimation method (e.g., $\widehat{\boldsymbol{\theta}}^{\mathrm{MLE}}$,
$\widehat{\boldsymbol{\theta}}^{\mathrm{NBE}}$,
$\widehat{\boldsymbol{\theta}}^{\mathrm{EOT}}$,
$\widehat{\boldsymbol{\theta}}^{\mathrm{AW}}$),
and $\boldsymbol{\theta}_0$ denotes the true parameter value. For two independent random vectors $\boldsymbol{X}$ and $\boldsymbol{Y}$ with respective distributions $\mu$ and $\nu$, we write $\mu \otimes \nu$ for their product measure, i.e., the joint distribution of $(\boldsymbol{X}, \boldsymbol{Y})$.

\section{Benchmark framework}\label{sec:framework}

\subsection{Overview of estimation methods}

The benchmark framework is defined by three components: (i) a collection of
data-generating processes, (ii) four point estimation procedures, and
(iii) evaluation metrics. The estimators are summarized in
Table~\ref{tab:methods_benchmark}.

\begin{table}[H]
\centering
\caption{\sf\small Overview of point estimation methods considered in the benchmark.}
\label{tab:methods_benchmark}
\begin{tabularx}{\textwidth}{|l|c|X|}
\hline
\textbf{Method} & \textbf{Category} & \textbf{Principle}  \\
\hline
MLE & Likelihood-based & Maximizes the log-likelihood when available. \\
\hline
NBE & Neural-based & Learns a Bayes point estimator from simulated parameter-data pairs. \\
\hline
EOT & Distance-based & Minimizes a Sinkhorn discrepancy between observed and simulated empirical distributions. \\
\hline
AW--NBE & Hybrid (Neural + EOT) & Refines the NBE by minimizing a regularized Sinkhorn criterion. \\
\hline
\end{tabularx}
\end{table}

More explicitly, the maximum log-likelihood estimator is defined, whenever the
likelihood $\mathcal{L}$ is available, by
\[
\widehat{\boldsymbol{\theta}}^{\mathrm{MLE}}
\in
\argmax_{\boldsymbol{\theta}\in\Theta}
\log \mathcal{L}_{model}(\boldsymbol{\theta}\mid\boldsymbol{x}).
\]
It is used as a likelihood-based benchmark under correct specification.

The NBE (\cite{sainsbury2024neural}) is trained offline on simulated pairs
$(\boldsymbol{\theta}^{(k)},\boldsymbol{X}^{(k)})$, for $k\geq 1$, where
$\boldsymbol{\theta}^{(k)}$ is drawn from a prior distribution
$\pi(\boldsymbol{\theta})$ and $\boldsymbol{X}^{(k)}$ is simulated from
$P_{\boldsymbol{\theta}^{(k)}}$. A neural network $f_{\psi}$ is fitted by
minimizing a squared-error loss, and the resulting estimator is
\[
\widehat{\boldsymbol{\theta}}^{\mathrm{NBE}}
=
f_{\widehat{\psi}}(\boldsymbol{x}).
\]
Under quadratic loss, the Bayes estimator is the posterior mean, since it 
minimizes the posterior expected loss
$\mathbb{E}\left[\|\boldsymbol{a}-\boldsymbol{\theta}\|^2 \mid \boldsymbol{X}\right]$
over all possible actions $\boldsymbol{a}$. Therefore, the NBE trained with a 
quadratic loss targets $\mathbb{E}[\boldsymbol{\theta}\mid\boldsymbol{X}]$ 
\citep[see e.g.,][]{casella2002statistical}.

The EOT estimator is defined as a minimum-distance
estimator between empirical distribution based on the Sinkhorn divergence. If $\widehat P_m$ denotes the
empirical distribution of the observed sample and
$\widehat P_{m_s,\boldsymbol{\theta}}$ the empirical distribution of a sample
simulated from $P_{\boldsymbol{\theta}}$, then
\[
\widehat{\boldsymbol{\theta}}^{\mathrm{EOT}}
\in
\argmin_{\boldsymbol{\theta}\in\Theta}
S_{\varepsilon}
\left(
\widehat P_m,
\widehat P_{m_s,\boldsymbol{\theta}}
\right),
\]
with $\varepsilon>0$ the entropic regularization parameter \cite[see e.g.,][]{Genevay2018}.
This estimator compares observed and simulated distributions directly, without
requiring likelihood evaluation.

Finally, the AW--NBE combines the previous
two likelihood-free strategies. It uses the NBE as an anchoring point and
refines it through a regularized Sinkhorn criterion:
\[
\widehat{\boldsymbol{\theta}}^{\mathrm{AW}}
\in
\argmin_{\boldsymbol{\theta}\in\Theta}
\left\{
S_{\varepsilon}
\left(
\widehat P_m,
\widehat P_{m_s,\boldsymbol{\theta}}
\right)
+
\lambda
\left\|
\boldsymbol{\theta}
-
\widehat{\boldsymbol{\theta}}^{\mathrm{NBE}}
\right\|_2^2
\right\},
\]
with $\lambda$ a tuning parameter.
The first term enforces distributional agreement between observed and simulated
data, while the second term keeps the refinement close to the neural estimate.
In this sense, AW--NBE is a Sinkhorn-based refinement of the NBE.

\subsection{Evaluation criteria}

The methods under study all return point estimates, but they are obtained from different criteria: 
the MLE maximizes a likelihood criterion and serves as a reference, the NBE finds a point estimator via a learned mapping based on a DeepSets architecture, whereas 
the EOT estimator is defined as a minimum distance estimator based on the Sinkhorn divergence 
\citep{cuturi2013sinkhorn, Genevay2018}, and the AW--NBE combines these two approaches. 
Consequently, the comparison cannot be restricted to a single criterion.  \\
Let $R$ denote the number of Monte Carlo replications and $m$ the sample size
within each replication. For replication $r=1,\ldots,R$, let
$\widehat{\boldsymbol{\theta}}_r$ be the estimate obtained from the $r$-th
dataset, and let $\boldsymbol{\theta}_0$ denote the reference parameter value.
We denote by $\|\cdot\|_2$ the Euclidean norm. Estimator performance is assessed
through the root mean squared error computed across Monte Carlo replications:
\begin{equation}
\label{eq:rmse}
\mathrm{RMSE}(\widehat{\boldsymbol{\theta}}) 
= \left(\frac{1}{R}\sum_{r=1}^R  
\|\widehat{\boldsymbol{\theta}}_r - \boldsymbol{\theta}_0\|_2^2\right)^{1/2}.
\end{equation}
To compare estimators more formally, we perform paired two-sided $t$-tests on 
the replication-wise squared-error losses \citep{casella2002statistical}. 
For each pair of estimators reported in the tables as Method A and Method B, 
we define the paired loss difference
\begin{equation}
\label{eq:paired_loss_difference}
D_r^{A,B}
=
\|\widehat{\boldsymbol{\theta}}^{A}_r-\boldsymbol{\theta}_0\|_2^2
-
\|\widehat{\boldsymbol{\theta}}^{B}_r-\boldsymbol{\theta}_0\|_2^2,
\qquad r=1,\ldots,R.
\end{equation}
The paired $t$-test is then applied to the sample 
$\{D_r^{A,B}\}_{r=1}^R$, with null hypothesis
\begin{equation}
\label{eq:paired_ttest_null}
H_0:\quad \mathbb{E}\!\left[D^{A,B}\right]=0.
\end{equation}
A positive $t$-statistic indicates that Method B has a lower mean squared error (MSE)
than Method A. The paired design accounts for the fact that the two estimators 
are evaluated on the same Monte Carlo replications, thereby reducing 
dataset-to-dataset variability in the comparison.
\subsection{Simulation design}

We consider three model families: a trivariate Gaussian model (light-tailed reference), a trivariate common shock Poisson model (discrete counts), and a trivariate Double Pareto model (heavy tails). Each is evaluated under correct specification and two misspecification scenarios. NBE and AW--NBE are trained on data simulated from the reference model with parameters drawn from a prior; MLE and EOT are evaluated directly on test data.

The numerical comparison is based on a Monte Carlo design. For each scenario, we generate $R = 100$ independent Monte Carlo replications, each consisting of a dataset of size $m = 50$. All estimators are evaluated on the same simulated datasets, yielding a paired design, and random seeds are fixed for reproducibility.\\
For each model, we follow the same structure: model specification, likelihood construction, scenario design, and presentation of results, followed by a discussion.\\[1ex]
The experiments were run on a standard laptop equipped with an 11th Gen Intel(R)
Core(TM) i5-1145G7 CPU at 2.60 GHz and 16 GB of RAM. No dedicated GPU
acceleration was used. Each model family required approximately 2.5 days of
computation, leading to a total runtime of about 7.5 days when the three model
families were run sequentially.

%\section{Results}\label{sec:results}

% Objectif :
% - Structurer résultats par phénomène

\section{Trivariate Gaussian model with AR(1)-type covariance}\label{sec:gauss}

\subsection{Model construction and properties}

We consider a trivariate Gaussian model with AR(1) covariance structure,
\begin{equation}
\label{eq:gaussian_ar1_model}
\boldsymbol{X} \sim \mathcal{N}_3\!\left(\mu\,\mathbf{1},\; \sigma^2 \Gamma(\rho)\right),
\qquad
\Gamma(\rho)_{ij} = \rho^{|i-j|},
\qquad
\boldsymbol{\theta} = (\mu,\sigma^2,\rho),
\end{equation}
with $\sigma^2>0$ and $\rho\in(-1,1)$.
AR(1)-type covariance structures are classical in time series and spatial
statistics, where correlations decay with distance or lag
\citep{brockwell1991time, cressie1993statistics}; they provide a convenient low-dimensional framework for assessing
the behavior of the competing estimators under both correct specification
and model misspecification, with controlled dependence through a single
correlation parameter $\rho$.
\subsection{Likelihood calculation}
For an observed sample $\boldsymbol{x}_1, \ldots, \boldsymbol{x}_m \in \mathbb{R}^3$,
the log-likelihood is given by
\begin{equation*}
 \log \mathcal{L}_G(\mu,\sigma^2,\rho \mid \boldsymbol{x})=-\frac{m}{2}\log\det\!\left(\sigma^2 \Gamma(\rho)\right)
  -\frac{1}{2}\sum_{i=1}^m
   (\boldsymbol{x}_i - \mu\mathbf{1})^\top
   \left(\sigma^2 \Gamma(\rho)\right)^{-1}
   (\boldsymbol{x}_i - \mu\mathbf{1}),
\end{equation*}
up to an additive constant independent of the parameters.
The MLE is obtained by numerical maximization of this criterion with respect
to $(\mu, \sigma^2, \rho)$.
\subsection{Scenario design}

The four estimators are compared under three scenarios.

\paragraph{Scenario 1: Well-specified baseline.}
Data are generated from Model~\eqref{eq:gaussian_ar1_model},
with $\boldsymbol{\theta}_0 = (1.0,\,2.0,\,0.6)$.
This scenario serves as a reference case under correct specification,
against which all estimators are evaluated.

\paragraph{Scenario 2: Inflated-variance contamination.}
Following Huber's gross-error model \citep{Huber1964, hampel1986robust},
observations are drawn from the mixture
\[
0.9\,\mathcal{N}_3(\mu\mathbf{1},\, \sigma^2 \Gamma(\rho))
+ 0.10\,\mathcal{N}_3(\mu\mathbf{1},\, 10\sigma^2 \Gamma(\rho)).
\]
This variance-inflation contamination generates atypical observations
with heavier tails, which are known to severely affect likelihood-based
procedures.

\paragraph{Scenario 3: Student-$t$ misspecification.}
Data are generated from a multivariate Student-$t$ distribution with
$\nu = 3$ degrees of freedom, same location and scale as the reference
Gaussian model \citep{lange1989robust, peel2000robust}.
This scenario evaluates the sensitivity of the estimators to tail
misspecification under a controlled departure from Gaussianity.

\subsection{Results}\label{subsubsec:results_Gaussian}
We first report the RMSE values and bootstrap confidence intervals for the
three Gaussian scenarios in Tables~\ref{tab:gauss_ar1_rmse_clean}--\ref{tab:gauss_ar1_rmse_t},
followed by the pairwise $t$-tests of $H_0$, as defined in
\eqref{eq:paired_ttest_null}, in Tables~\ref{tab:gauss_ar1_ttest_clean}--\ref{tab:gauss_ar1_ttest_t}.

\begin{table}[H]
\centering
\caption{\sf\small Trivariate Gaussian AR(1), clean scenario.
RMSE$(\widehat{\boldsymbol{\theta}})$ and 95\% bootstrap confidence interval, $R=100$, $m=50$.}
\label{tab:gauss_ar1_rmse_clean}
\begin{tabular}{lccc}
\toprule
Method & RMSE$(\widehat{\boldsymbol{\theta}})$ & 95\% CI (bootstrap) & $\Delta_{\text{best}}$ \\
\midrule
$\boldsymbol{\mathrm{MLE}}$ & $\boldsymbol{0.3272}$ & $\boldsymbol{[0.2970,\;0.3562]}$ & --- \\
EOT     & 0.5205 & [0.4854,\;0.5571] & $+59.1\%$ \\
NBE     & 0.3707 & [0.3271,\;0.4131] & $+13.3\%$ \\
AW--NBE & 0.4102 & [0.3764,\;0.4399] & $+25.4\%$ \\
\bottomrule
\end{tabular}
\end{table}

\begin{table}[H]
\centering
\caption{\sf\small Trivariate Gaussian AR(1), 10\% inflated-variance contamination.
RMSE$(\widehat{\boldsymbol{\theta}})$ and 95\% bootstrap confidence interval, $R=100$, $m=50$.}
\label{tab:gauss_ar1_rmse_contam}
\begin{tabular}{lccc}
\toprule
Method & RMSE$(\widehat{\boldsymbol{\theta}})$ & 95\% CI (bootstrap) & $\Delta_{\text{best}}$ \\
\midrule
MLE     & 1.8044 & [1.6430,\;1.9722] & $+180.8\%$ \\
$\boldsymbol{\mathrm{EOT}}$ & $\boldsymbol{0.6427}$ & $\boldsymbol{[0.5210,\;0.7702]}$ & --- \\
NBE     & 1.3640 & [1.2364,\;1.4838] & $+112.2\%$ \\
AW--NBE & 0.9913 & [0.8752,\;1.1119] & $+54.2\%$ \\
\bottomrule
\end{tabular}
\end{table}

\begin{table}[H]
\centering
\caption{\sf\small Trivariate Student-$t$ ($\nu=3$) misspecification.
RMSE$(\widehat{\boldsymbol{\theta}})$ and 95\% bootstrap confidence interval, $R=100$, $m=50$.}
\label{tab:gauss_ar1_rmse_t}
\begin{tabular}{lccc}
\toprule
Method & RMSE$(\widehat{\boldsymbol{\theta}})$ & 95\% CI (bootstrap) & $\Delta_{\text{best}}$ \\
\midrule
MLE     & 2.4028 & [2.2778,\;2.5194] & $+132.7\%$ \\
$\boldsymbol{\mathrm{EOT}}$ & $\boldsymbol{1.0326}$ & $\boldsymbol{[0.9244,\;1.1342]}$ & --- \\
NBE     & 1.9045 & [1.8020,\;2.0035] & $+84.4\%$ \\
AW--NBE & 1.5257 & [1.4240,\;1.6284] & $+47.8\%$ \\
\bottomrule
\end{tabular}
\end{table}
\begin{table}[H]
\centering
\caption{\sf\small Paired $t$-tests with $H_0$ defined in \eqref{eq:paired_ttest_null}, clean scenario.
Column "Winner" indicates the 'best' method (in terms of MSE) between A and B.
}
\label{tab:gauss_ar1_ttest_clean}
\begin{tabular}{llccl}
\toprule
Method A & Method B & $t$ stat. & $p$-value & Winner \\
\midrule
MLE & EOT    & $-7.639$ & $<0.001^{***}$ & A (MLE) \\
MLE & NBE    & $-3.514$ & $<0.001^{***}$ & A (MLE) \\
MLE & AW--NBE & $-4.823$ & $<0.001^{***}$ & A (MLE) \\
EOT & NBE    & $\phantom{-}4.667$ & $<0.001^{***}$ & B (NBE) \\
EOT & AW--NBE & $\phantom{-}10.383$ & $<0.001^{***}$ & B (AW--NBE) \\
NBE & AW--NBE & $-1.591$ & $0.114$ & A (NBE) \\
\bottomrule
\end{tabular}
\end{table}

\begin{table}[H]
\centering
\caption{\sf\small Paired $t$-tests with $H_0$ defined in \eqref{eq:paired_ttest_null}, 10\% contamination scenario. Column "Winner" indicates the 'best' method between A and B.
}
\label{tab:gauss_ar1_ttest_contam}
\begin{tabular}{llccl}
\toprule
Method A & Method B & $t$ stat. & $p$-value & Winner \\
\midrule
MLE & EOT    & $\phantom{-}9.775$ & $<0.001^{***}$ & B (EOT) \\
MLE & NBE    & $\phantom{-}7.549$ & $<0.001^{***}$ & B (NBE) \\
MLE & AW--NBE & $\phantom{-}9.581$ & $<0.001^{***}$ & B (AW--NBE) \\
EOT & NBE    & $-11.893$ & $<0.001^{***}$ & A (EOT) \\
EOT & AW--NBE & $-8.874$  & $<0.001^{***}$ & A (EOT) \\
NBE & AW--NBE & $\phantom{-}14.171$ & $<0.001^{***}$ & B (AW--NBE) \\
\bottomrule
\end{tabular}
\end{table}

\begin{table}[H]
\centering
\caption{\sf\small Paired $t$-tests with $H_0$ defined in \eqref{eq:paired_ttest_null}, Student-$t$ ($\nu=3$) misspecification scenario. Column "Winner" indicates the 'best' method between A and B.
}
\label{tab:gauss_ar1_ttest_t}
\begin{tabular}{llccl}
\toprule
Method A & Method B & $t$ stat. & $p$-value & Winner \\
\midrule
MLE & EOT    & $\phantom{-}17.692$ & $<0.001^{***}$ & B (EOT) \\
MLE & NBE    & $\phantom{-}11.122$ & $<0.001^{***}$ &   B (NBE) \\
MLE & AW--NBE & $\phantom{-}15.601$ & $<0.001^{***}$ & B (AW--NBE) \\
EOT & NBE    & $-24.045$ & $<0.001^{***}$ & A (EOT) \\
EOT & AW--NBE & $-18.824$ & $<0.001^{***}$ & A (EOT) \\
NBE & AW--NBE & $\phantom{-}28.573$ & $<0.001^{***}$ & B (AW--NBE) \\
\bottomrule
\end{tabular}
\end{table}

\paragraph{Discussions}~\\[2ex]
\textit{\textbf{On the scenarios:}}
\begin{enumerate}
\item {\bf Clean scenario.}
The MLE achieves the smallest
RMSE$(\widehat{\boldsymbol{\theta}}) = 0.327$, consistent with its asymptotic efficiency.
All three likelihood-free estimators are significantly outperformed by the MLE. 
We also notice that NBE and AW--NBE are close.
EOT is the worst performer in this scenario.
\item {\bf Contamination scenario.}
The MLE deteriorates severely, with its RMSE rising from $0.327$ to $1.804$ under contamination, confirming its high sensitivity to heavy-tailed perturbations.
Moreover, its confidence interval is wide and nearly symmetric, with the RMSE
lying about $0.16$ units from each bound. This indicates substantial replication-to-replication variability in the estimation error.
EOT is the most robust estimator, while AW--NBE substantially reduces
the error of the baseline NBE.
This 27\% reduction (compared to the NBE, so not $\Delta$) constitutes the central empirical argument in favor of AW--NBE:
the Wasserstein-based refinement attenuates the sensitivity to atypical observations
inherited by the NBE from its squared-error training loss.
\item {\bf Student-$t$ misspecification.} The same hierarchy as in the contamination scenario is preserved.
The MLE degrades most severely, EOT remains the most robust, and AW--NBE improves over NBE by approximately 20\%.
\end{enumerate}

\textit{\textbf{Overall:}}\\[1ex]
Across the three scenarios, no single estimator dominates uniformly.
The MLE performs best under correct specification (see Tables~\ref{tab:gauss_ar1_ttest_clean}), as expected for likelihood-based estimation in these regular settings, but sensitive under deviation.
EOT provides the strongest robustness under misspecification, but its performance under correct specification is consistently the weakest among all methods considered.
AW--NBE occupies an intermediate position: it matches NBE under correct specification and consistently improves upon it under misspecification,
offering a principled robustness--efficiency trade-off within the class of neural estimators.

\section{The trivariate common shock Poisson distribution}
\label{sec:trivariate_poisson}

\subsection{Model construction and properties}
Following \cite{Inouye2017}, we use the common shock construction \citep{mckendrick1925applications,campbell1934poisson,teicher1954multivariate}. Let $Y_1, Y_2, Y_3, Z$ be mutually independent Poisson variables with rates $\boldsymbol{\kappa}=(\kappa_0,\kappa_1,\kappa_2,\kappa_3)\in \mathbb{R}_+^4$. The trivariate vector $\boldsymbol{X} = (X_1, X_2, X_3)\in \mathbb{Z}^3_+$ is defined by
\begin{equation}
\label{eq:common_shock_poisson}
X_i = Y_i + Z, \qquad i=1,2,3.
\end{equation}
By the reproductive property, $X_i \sim \mathrm{Poisson}(\kappa_i + \kappa_0)$ and $\mathrm{Cov}(X_i, X_j) = \kappa_0$ for $i \neq j$, restricting the model to positive dependence.

\subsection{Likelihood calculation}

Conditioning on the common shock $Z=z$ and marginalizing over its possible
values yields the following log-likelihood for an independent sample \citep{teicher1954multivariate}:
\begin{equation}
\label{eq:common_shock_pmf}
\mathcal{\log L_P(\boldsymbol{\kappa}|\boldsymbol{x})}
=
-m\sum_{i=0}^3 \kappa_i
+
\sum_{j=1}^m
\log\left[
\sum_{z=0}^{\min(x_{j1},x_{j2},x_{j3})}
\frac{\kappa_0^z}{z!}
\prod_{i=1}^3
\frac{\kappa_i^{\,x_{ji}-z}}{(x_{ji}-z)!}
\right].
\end{equation}
The MLE is obtained by maximizing this log-likelihood \eqref{eq:common_shock_pmf} over $\boldsymbol{\kappa}\in\mathbb{R}_+^4$.

\subsection{Scenario design}
The four estimators are compared under three scenarios.

\paragraph{Scenario 1: Well-specified baseline.}
Data are generated from Model~\eqref{eq:common_shock_poisson}
under correct specification.

\paragraph{Scenario 2: Strong Poisson--Gamma overdispersion ($r=5$).} This construction is motivated by the classical Poisson--Gamma representation of the negative binomial distribution, widely used for overdispersed count data
\citep{Lawless1987,Cameron1998,Inouye2017}. 
Fixed rates are replaced by Gamma-distributed intensities:
\[
X_i \mid G_0,G_i \sim \mathrm{Poisson}(G_0+G_i),
\qquad
G_j \sim \Gamma\!\left(r, \tfrac{r}{\kappa_j}\right), \quad j = 0,1,2,3.
\]
This Poisson--Gamma construction preserves $\mathbb{E}(X_i)=\kappa_0+\kappa_i$ and inflates the marginal variance to
\[
\mathrm{Var}(X_i) = \kappa_0+\kappa_i + \frac{\kappa_0^2+\kappa_i^2}{r},
\]
giving a relative standard deviation (RSD) of approximately $57.7\%$ for $r=5$.

\paragraph{Scenario 3: Mild Poisson--Gamma overdispersion ($r=20$).}
Same construction, leading to RSD $\approx 50.0\%$ for $r=20$. 

\subsection{Results}
As in Section~\ref{subsubsec:results_Gaussian}, Tables~\ref{tab:pois_cs_rmse_clean}--\ref{tab:pois_cs_rmse_poisgamma_light}
report RMSE$(\widehat{\boldsymbol{\theta}})$ with bootstrap confidence intervals,
and Tables~\ref{tab:pois_cs_ttest_clean}--\ref{tab:pois_cs_ttest_poisgamma_light}
the corresponding pairwise paired $t$-tests.
\begin{table}[H]
\centering
\caption{\sf\small Trivariate common shock Poisson, clean scenario
($\kappa_0=1.5$, $\kappa_i=3.0$).
RMSE$(\widehat{\boldsymbol{\theta}})$ and 95\% bootstrap confidence interval, $R=100$, $m=50$.}
\label{tab:pois_cs_rmse_clean}
\begin{tabular}{lccc}
\toprule
Method & RMSE$(\widehat{\boldsymbol{\theta}})$ & 95\% CI (bootstrap) & $\Delta_{\text{best}}$ \\
\midrule
$\boldsymbol{\mathrm{MLE}}$    & $\boldsymbol{0.5673}$ & $\boldsymbol{[0.4945,\; 0.6384]}$ & --- \\
EOT    & 0.7378 & [0.6602,\; 0.8299] & $+30.1\%$ \\
NBE    & 0.6549 & [0.5900,\; 0.7187] & $+15.4\%$ \\
AW--NBE & 0.6460 & [0.5737,\; 0.7071] & $+13.9\%$ \\
\bottomrule
\end{tabular}
\end{table}

\begin{table}[H]
\centering
\caption{\sf\small Trivariate common shock Poisson, Poisson--Gamma overdispersion ($r=5$).
RMSE$(\widehat{\boldsymbol{\theta}})$ and 95\% bootstrap confidence interval, $R=100$, $m=50$.}
\label{tab:pois_cs_rmse_poisgamma}
\begin{tabular}{lccc}
\toprule
Method & RMSE$(\widehat{\boldsymbol{\theta}})$ & 95\% CI (bootstrap) & $\Delta_{\text{best}}$ \\
\midrule
MLE    & 1.9425 & [1.9063,\; 1.9829] & $+38.4\%$ \\
$\boldsymbol{\mathrm{EOT}}$ & $\boldsymbol{1.4034}$ & $\boldsymbol{[1.3450,\; 1.4630]}$ & --- \\
NBE    & 2.1654 & [2.1175,\; 2.2139] & $+54.3\%$ \\
AW--NBE & 1.7987 & [1.7689,\; 1.8257] & $+28.2\%$ \\
\bottomrule
\end{tabular}
\end{table}

\begin{table}[H]
\centering
\caption{\sf\small Trivariate common shock Poisson, mild Poisson--Gamma overdispersion
($r=20$, variance/mean $= 1.125$).
RMSE$(\widehat{\boldsymbol{\theta}})$ and 95\% bootstrap confidence interval, $R=100$, $m=50$.}
\label{tab:pois_cs_rmse_poisgamma_light}
\begin{tabular}{lccc}
\toprule
Method & RMSE$(\widehat{\boldsymbol{\theta}})$ & 95\% CI (bootstrap) & $\Delta_{\text{best}}$ \\
\midrule
MLE    & 1.8501 & [1.7989,\; 1.9039] & $+36.6\%$ \\
$\boldsymbol{\mathrm{EOT}}$    & $\boldsymbol{1.3541}$ & $\boldsymbol{[1.2859,\; 1.4137]}$ & --- \\
NBE    & 1.6923 & [1.6340,\; 1.7473] & $+25.0\%$ \\
AW--NBE & 1.6594 & [1.6128,\; 1.7162] & $+22.5\%$ \\
\bottomrule
\end{tabular}
\end{table}

\begin{table}[H]
\centering
\caption{\sf\small Paired $t$-tests with $H_0$ defined in \eqref{eq:paired_ttest_null}, clean scenario.
Column "Winner" indicates the 'best' method (MSE criterion) between A and B.
}
\label{tab:pois_cs_ttest_clean}
\begin{tabular}{llccl}
\toprule
Method A & Method B & $t$ stat. & $p$-value & Winner \\
\midrule
MLE & EOT    & $-3.527$ & $<0.001^{***}$ & A (MLE) \\
MLE & NBE    & $-2.874$ & $0.005^{**}$   & A (MLE) \\
MLE & AW--NBE & $-2.593$ & $0.011^{*}$    & A (MLE) \\
EOT & NBE    & $\phantom{-}1.633$ & $0.106$ & B (NBE) \\
EOT & AW--NBE & $\phantom{-}1.849$ & $0.067$ & B (AW--NBE) \\
NBE & AW--NBE & $\phantom{-}0.807$ & $0.422$ & B (AW--NBE) \\
\bottomrule
\end{tabular}
\end{table}

\begin{table}[H]
\centering
\caption{\sf\small Paired $t$-tests with $H_0$ defined in \eqref{eq:paired_ttest_null}, Poisson--Gamma overdispersion ($r=5$).
Column "Winner" indicates the 'best' method between A and B.}
\label{tab:pois_cs_ttest_poisgamma}
\begin{tabular}{llccl}
\toprule
Method A & Method B & $t$ stat. & $p$-value & Winner \\
\midrule
MLE & EOT    & $\phantom{-}21.588$ & $<0.001^{***}$ & B (EOT) \\
MLE & NBE    & $-13.807$           & $<0.001^{***}$ & A (MLE) \\
MLE & AW--NBE & $\phantom{-}11.446$ & $<0.001^{***}$ & B (AW--NBE) \\
EOT & NBE    & $-25.016$           & $<0.001^{***}$ & A (EOT) \\
EOT & AW--NBE & $-14.887$           & $<0.001^{***}$ & A (EOT) \\
NBE & AW--NBE & $\phantom{-}17.910$ & $<0.001^{***}$ & B (AW--NBE) \\
\bottomrule
\end{tabular}
\end{table}

\begin{table}[H]
\centering
\caption{\sf\small Paired $t$-tests with $H_0$ defined in \eqref{eq:paired_ttest_null}, mild Poisson--Gamma overdispersion ($r=20$).
Column "Winner" indicates the 'best' method between A and B.}
\label{tab:pois_cs_ttest_poisgamma_light}
\begin{tabular}{llccl}
\toprule
Method A & Method B & $t$ stat. & $p$-value & Winner \\
\midrule
MLE & EOT    & $\phantom{-}18.005$ & $<0.001^{***}$ & B (EOT) \\
MLE & NBE    & $\phantom{-}7.683$  & $<0.001^{***}$ & B (NBE) \\
MLE & AW--NBE & $\phantom{-}10.017$ & $<0.001^{***}$ & B (AW--NBE) \\
EOT & NBE    & $-10.106$           & $<0.001^{***}$ & A (EOT) \\
EOT & AW--NBE & $-9.858$            & $<0.001^{***}$ & A (EOT) \\
NBE & AW--NBE & $\phantom{-}3.728$  & $<0.001^{***}$ & B (AW--NBE) \\
\bottomrule
\end{tabular}
\end{table}

\paragraph{Discussions}~\\[2ex]
\textit{\textbf{On the scenarios:}}
\begin{enumerate}
\item {\bf Clean scenario.}
The MLE achieves the smallest
RMSE$(\widehat{\boldsymbol{\theta}}) = 0.567$, as expected from likelihood-based
estimation in a well-specified model.
All three likelihood-free estimators are significantly outperformed by the MLE
(Table~\ref{tab:pois_cs_rmse_clean}).
The gap remains moderate for NBE ($\Delta = +15.4\%$) and AW--NBE
($\Delta = +13.9\%$), while EOT is the worst performer ($\Delta = +30.1\%$),
reflecting the efficiency loss of the transport criterion under correct
specification.
The difference between AW--NBE and NBE does not reach significance ($p = 0.422$,
Table~\ref{tab:pois_cs_ttest_clean}), though AW--NBE shows a slightly lower RMSE.

\item {\bf Poisson--Gamma overdispersion ($r=5$).}
The MLE deteriorates severely, with its RMSE rising from $0.567$ to $1.943$ under overdispersion ($\Delta = +38.4\%$ above EOT, Table~\ref{tab:pois_cs_rmse_poisgamma}), confirming its sensitivity to extra-Poisson variability. A notable inversion occurs: NBE becomes the worst estimator of all ($\Delta = +54.3\%$), even significantly worse than the MLE (Table~\ref{tab:pois_cs_ttest_poisgamma}), suggesting that the network has
implicitly learned a Poisson-specific representation during training and fails
to generalize under distributional shift.
EOT is by far the best-performing method, significantly outperforming every
alternative.
AW--NBE occupies an intermediate position ($\Delta = +28.2\%$): its RMSE 
drops to $1.799$ compared to $2.165$ for NBE (a reduction of approximately 
$17\%$), and it significantly outperforms the MLE, showing that the Wasserstein-based refinement partially corrects the sensitivity inherited from the squared-error training objective.

\item {\bf Mild Poisson--Gamma overdispersion ($r=20$).}
EOT remains the best performer
(RMSE$(\widehat{\boldsymbol{\theta}})= 1.354$), significantly ahead of all
competitors (Table~\ref{tab:pois_cs_rmse_poisgamma_light}).
The MLE and NBE lag behind by $\Delta = +36.6\%$ and $\Delta = +25.0\%$
respectively, while AW--NBE narrows the gap to $\Delta = +22.5\%$.
Crucially, both NBE and AW--NBE now improve upon the MLE
(Table~\ref{tab:pois_cs_ttest_poisgamma_light}), indicating that even a
moderate departure from the Poisson assumption is sufficient to erode the
MLE's advantage.
The fragility of NBE observed under strong overdispersion is attenuated here,
yet AW--NBE still significantly outperforms it, confirming that the adaptive
Wasserstein refinement provides a consistent, if moderate, gain across the
range of overdispersion severity considered.
\end{enumerate}
\textit{\textbf{Overall:}}\\[1ex]
Across the three scenarios, the results mirror and extend the conclusions
drawn for the Gaussian model.
The MLE is the best under correct specification but degrades significantly
under distributional misspecification.
EOT provides the strongest robustness at the cost of a meaningful efficiency
loss under correct specification.
AW--NBE consistently improves upon NBE under misspecification while remaining
comparable to it under correct specification.
A distinctive feature of this experiment is the severe degradation of NBE
under strong overdispersion ($r=5$), where it is outperformed even by the
misspecified MLE.
\section{Trivariate Double Pareto}\label{sec:doublePareto}

\subsection{Model construction and properties}

To study the behavior of the competing estimators in a heavy-tailed setting,
we build a trivariate model from two independent univariate components.
Let $U$ and $V$ be independent random variables, each following a
\emph{univariate} symmetric Double Pareto distribution, denoted
$U \sim \mathrm{DP}(\alpha_u)$ and $V \sim \mathrm{DP}(\alpha_v)$,
in the sense of \cite{Al-Athari2011}. The univariate $\mathrm{DP}(\alpha)$
distribution has density
\begin{equation}
\label{eq:dp_density}
f(x;\alpha) = \frac{\alpha}{2}(1+|x|)^{-(\alpha+1)},
\quad x \in \mathbb{R}, \quad \alpha > 1.
\end{equation}
The trivariate observation vector is then defined by the deterministic
construction
\[
(X_1, X_2, X_3) = (U,\, V,\, U+V),
\]
so that the distribution of $(X_1, X_2, X_3)$ is entirely characterized by
the two scalar parameters $\boldsymbol{\theta}=(\alpha_u,\alpha_v)$. We do
not assume a separate multivariate Double Pareto family; the multivariate
structure arises solely from the linear constraint $X_3=X_1+X_2$, together
with the independence of $U$ and $V$. This constraint induces dependence
between $X_3$ and the first two components, while $X_1$ and $X_2$ remain
independent. The true parameter is
$\boldsymbol{\theta}_0  = (2.2,\, 4.0)$.
\subsection{Likelihood calculation}

Since $X_3 = U+V$ is a deterministic function of $(U,V)$,
the joint distribution of $(X_1,X_2,X_3)$ is singular with respect
to the Lebesgue measure on $\mathbb{R}^3$. The likelihood is therefore
expressed in terms of the bivariate marginal of $(U,V) = (X_1, X_2)$
only, since the third component carries no additional information
beyond $(U,V)$. For an observed sample
$\boldsymbol{x}_1,\ldots,\boldsymbol{x}_m$, where each
$\boldsymbol{x}_i = (u_i, v_i, u_i+v_i)$, the independence of $U$
and $V$ yields the decomposed log-likelihood
\begin{equation}
\label{eq:dp_loglik}
\log \mathcal{L}_{\mathrm{DP}}(\alpha_u,\alpha_v \mid \boldsymbol{x})
=
\sum_{i=1}^m \log f(u_i;\alpha_u)
+
\sum_{i=1}^m \log f(v_i;\alpha_v),
\end{equation}
where $f(\,\cdot\,;\alpha)$ is given in \eqref{eq:dp_density}.
Each marginal log-likelihood is maximized separately, yielding
closed-form MLEs:
\[
\widehat{\alpha}_u
= \frac{m}{\displaystyle\sum_{i=1}^m \log(1+|u_i|)},
\qquad
\widehat{\alpha}_v
= \frac{m}{\displaystyle\sum_{i=1}^m \log(1+|v_i|)}.
\]
\subsection{Scenario design}

We compare the four estimators under three scenarios.

\paragraph{Scenario 1: Well-specified baseline.}
Data are generated from the model class used for inference,
with independent $U \sim \mathrm{DP}(\alpha_u)$ and
$V \sim \mathrm{DP}(\alpha_v)$.
This scenario serves as a reference case under correct specification.

\paragraph{Scenario 2: $\varepsilon$-contamination.}
We consider a 10\% contamination, where each observation is drawn from
$$
0.9\bigl[\mathrm{DP}(\alpha_u)\otimes\mathrm{DP}(\alpha_v)\bigr]
+
0.1\bigl[\mathrm{DP}(1.6)\otimes\mathrm{DP}(2.8)\bigr],
$$
corresponding to a heavier-tailed contaminating distribution.
This follows the classical Huber contamination paradigm
\citep{Huber1964, Huber1981}.

\paragraph{Scenario 3: Dependence misspecification.}
The marginal distributions remain $\mathrm{DP}(\alpha_u)$ and
$\mathrm{DP}(\alpha_v)$, but $(U,V)$ are coupled via a
Student-$t$ copula with correlation $\rho = 0.5$ and $\nu = 4$
degrees of freedom, violating the independence assumption
underlying the model \citep{ColesHeffernanTawn1999}.

\subsection{Results}
The results for the double Pareto are presented in the subsequent tables. Tables~\ref{tab:dp_rmse_clean}--\ref{tab:dp_rmse_t} report RMSE$(\widehat{\boldsymbol{\theta}})$ with bootstrap confidence intervals, and Tables~\ref{tab:dp_ttest_clean}--\ref{tab:dp_ttest_t} the corresponding pairwise paired $t$-tests.

\begin{table}[H]
\centering
\caption{\sf\small Trivariate double Pareto, clean scenario
($\alpha_u=2.2$, $\alpha_v=4.0$).
RMSE$(\widehat{\boldsymbol{\theta}})$ and 95\% bootstrap confidence interval,
$R=100$, $m=50$.}
\label{tab:dp_rmse_clean}
\begin{tabular}{lccc}
\toprule
Method & RMSE$(\widehat{\boldsymbol{\theta}})$ & 95\% CI (bootstrap) & $\Delta_{\text{best}}$ \\
\midrule
$\boldsymbol{\mathrm{MLE}}$    & $\boldsymbol{0.1746}$ & $\boldsymbol{[0.1514,\; 0.2307]}$ & --- \\
EOT    & 1.1742 & [1.0144,\; 1.5028] & $+572.5\%$ \\
NBE    & 0.5740 & [0.5715,\; 0.5789] & $+228.7\%$ \\
AW--NBE & 0.5594 & [0.5560,\; 0.5661] & $+220.4\%$ \\
\bottomrule
\end{tabular}
\end{table}

\begin{table}[H]
\centering
\caption{\sf\small Trivariate double Pareto, 10\% heavy-tail contamination.
RMSE$(\widehat{\boldsymbol{\theta}})$ and 95\% bootstrap confidence interval,
$R=100$, $m=50$.}
\label{tab:dp_rmse_contam}
\begin{tabular}{lccc}
\toprule
Method & RMSE$(\widehat{\boldsymbol{\theta}})$ & 95\% CI (bootstrap) & $\Delta_{\text{best}}$ \\
\midrule
MLE    & 0.7884 & [0.7815,\; 0.7919] & $+33.3\%$ \\
$\boldsymbol{\mathrm{EOT}}$   & $\boldsymbol{0.5916}$ & $\boldsymbol{[0.5144,\; 0.7515]}$ & --- \\
NBE    & 0.7173 & [0.7113,\; 0.7202] & $+21.2\%$ \\
AW--NBE & 0.6108 & [0.6104,\; 0.6122] & $+3.2\%$ \\
\bottomrule
\end{tabular}
\end{table}

\begin{table}[H]
\centering
\caption{\sf\small Trivariate double Pareto, Student-$t$ copula dependence
misspecification ($\rho=0.5$, $\nu=4$).
RMSE$(\widehat{\boldsymbol{\theta}})$ and 95\% bootstrap confidence interval,
$R=100$, $m=50$.}
\label{tab:dp_rmse_t}
\begin{tabular}{lccc}
\toprule
Method & RMSE$(\widehat{\boldsymbol{\theta}})$ & 95\% CI (bootstrap) & $\Delta_{\text{best}}$ \\
\midrule
MLE    & 0.6448 & [0.6414,\; 0.6465] & $+110.8\%$ \\
EOT    & 0.5943 & [0.5801,\; 0.6287] & $+94.3\%$ \\
NBE    & 0.5883 & [0.5861,\; 0.5893] & $+92.4\%$ \\
$\textbf{AW--NBE}$ & $\boldsymbol{0.3058}$ & $\boldsymbol{[0.2929,\; 0.3417]}$ & --- \\
\bottomrule
\end{tabular}
\end{table}

\begin{table}[H]
\centering
\caption{\sf\small Paired $t$-tests with $H_0$ defined in \eqref{eq:paired_ttest_null}, clean scenario.
Column "Winner" indicates the 'best' method between A and B.
}
\label{tab:dp_ttest_clean}
\begin{tabular}{llccl}
\toprule
Method A & Method B & $t$ stat. & $p$-value & Winner \\
\midrule
MLE & EOT    & $-6.588$ & $<0.001^{***}$ & A (MLE) \\
MLE & NBE & $-3.021$ & $0.003^{**}$ & A (MLE) \\
MLE & AW--NBE & $-3.813$ & $<0.001^{***}$ & A (MLE) \\
EOT & NBE    & $\phantom{-}6.496$ & $<0.001^{***}$ & B (NBE) \\
EOT & AW--NBE & $\phantom{-}6.030$ & $<0.001^{***}$ & B (AW--NBE) \\
NBE & AW--NBE & $\phantom{-}7.636$ & $<0.001^{***}$ & B (AW--NBE) \\
\bottomrule
\end{tabular}
\end{table}

\begin{table}[H]
\centering
\caption{\sf\small Paired $t$-tests with $H_0$ defined \eqref{eq:paired_ttest_null}, 10\% contamination scenario. Column "Winner" indicates the 'best' method between A and B.
}
\label{tab:dp_ttest_contam}
\begin{tabular}{llccl}
\toprule
Method A & Method B & $t$ stat. & $p$-value & Winner \\
\midrule
MLE & EOT    & $\phantom{-}5.403$ & $<0.001^{***}$ & B (EOT) \\
MLE & NBE    & $\phantom{-}3.359$ & $0.001^{**}$ & B (NBE) \\
MLE & AW--NBE & $\phantom{-}1.697$ & $0.093$ & B (AW--NBE) \\
EOT & NBE    & $-5.443$ & $<0.001^{***}$ & A (EOT) \\
EOT & AW--NBE & $-5.285$ & $<0.001^{***}$ & A (EOT) \\
NBE & AW--NBE & $\phantom{-}3.696$ & $<0.001^{***}$ & B (AW--NBE) \\
\bottomrule
\end{tabular}
\end{table}

\begin{table}[H]
\centering
\caption{\sf\small Paired $t$-tests with $H_0$ defined \eqref{eq:paired_ttest_null}, Student-$t$ copula misspecification scenario.
Column "Winner" indicates the 'best' method between A and B.
}
\label{tab:dp_ttest_t}
\begin{tabular}{llccl}
\toprule
Method A & Method B & $t$ stat. & $p$-value & Winner \\
\midrule
MLE & EOT    & $\phantom{-}7.616$ & $<0.001^{***}$ & B (EOT) \\
MLE & NBE    & $\phantom{-}3.930$ & $<0.001^{***}$ & B (NBE) \\
MLE & AW--NBE & $\phantom{-}3.027$ & $0.003^{**}$ & B (AW--NBE) \\
EOT & NBE    & $\phantom{-}8.451$ & $<0.001^{***}$ & B (NBE) \\
EOT & AW--NBE & $\phantom{-}7.542$ & $<0.001^{***}$ & B (AW--NBE) \\
NBE & AW--NBE & $\phantom{-}7.985$ & $<0.001^{***}$ & B (AW--NBE) \\
\bottomrule
\end{tabular}
\end{table}

\paragraph{Discussions}~\\[2ex]
\textit{\textbf{On the scenarios:}}

\begin{enumerate}
\item {\bf Clean scenario.}
The MLE achieves the smallest RMSE$(\widehat{\boldsymbol{\theta}})=0.175$,
consistent with its asymptotic efficiency and mirroring the results observed
in the Gaussian and Poisson settings.
All three likelihood-free estimators are significantly outperformed by the MLE
(Table~\ref{tab:dp_rmse_clean}).
EOT is by far the worst performer ($\Delta = +572.5\%$), with the widest
confidence interval in the benchmark.
The interval is also clearly asymmetric around the RMSE estimate: the upper
deviation is about $0.33$, compared with $0.16$ on the lower side, suggesting
that the bootstrap distribution of the RMSE has a heavier upper tail,
consistent with occasional replications in which the Sinkhorn criterion is
strongly affected by extreme observations.

By contrast, NBE and AW--NBE display extremely narrow confidence intervals,
with deviations below $0.01$ on both sides, yet their gaps remain large
($\Delta = +228.7\%$ and $\Delta = +220.4\%$ respectively).
These intervals should not be interpreted as evidence of high accuracy, but
rather as evidence of highly concentrated errors across replications,
consistent with a low-variance but biased behavior of the neural estimators
in this heavy-tailed setting.
Among the likelihood-free estimators, AW--NBE achieves a small but
statistically significant improvement over NBE
(Table~\ref{tab:dp_ttest_clean}), suggesting that the transport-based
refinement brings a marginal gain even under correct specification.

The MLE deteriorates significantly, with its RMSE increasing by a factor of 4.5 relative to the clean case ($\Delta = +33.3\%$ above EOT, Table~\ref{tab:dp_rmse_contam}), confirming its sensitivity to heavy-tailed outliers. Unlike in the Gaussian contamination scenario, the MLE has an extremely
narrow confidence interval here, suggesting that its deterioration is not driven by unstable replications, but by a concentrated displacement toward a misspecified target.
EOT achieves the lowest RMSE and significantly outperforms all competitors (Table~\ref{tab:dp_ttest_contam}).
AW--NBE closely follows ($\Delta = +3.2\%$) and significantly improves upon
NBE ($\Delta = +21.2\%$), demonstrating that the Wasserstein-based refinement
partially corrects the sensitivity of the neural estimator to atypical
observations.
Notably, the improvement of AW--NBE over MLE does not reach statistical
significance ($p = 0.093$), indicating a borderline advantage only in this
scenario.

\item {\bf Student-$t$ copula misspecification.}
AW--NBE achieves by far the smallest
RMSE$(\widehat{\boldsymbol{\theta}})=0.306$, approximately half that of NBE,
EOT, and MLE (Table~\ref{tab:dp_rmse_t}), with $\Delta = +92.4\%$,
$\Delta = +94.3\%$, and $\Delta = +110.8\%$ for the three competing methods
respectively.
All pairwise differences involving AW--NBE are statistically significant
(Table~\ref{tab:dp_ttest_t}).
The near-identical RMSEs of EOT and NBE are noteworthy, though NBE is
significantly better than EOT ($p < 0.001$).
\end{enumerate}

\textit{\textbf{Overall:}}\\[1ex]
Across the three scenarios, the Double Pareto results refine and extend
the conclusions drawn for the Gaussian and Poisson settings.
The MLE wins under correct specification, consistent with its asymptotic
efficiency, but deteriorates substantially under both contamination and dependence misspecification.
EOT provides the strongest robustness to contamination but remains
the worst performer under correct specification.
NBE occupies an intermediate position: competitive under correct specification
and dependence misspecification, it improves upon the MLE under contamination
(RMSE$(\widehat{\boldsymbol{\theta}})= 0.717$ vs.\ $0.788$).
AW--NBE provides the most balanced behavior among the likelihood-free estimators considered here:
it achieves a small but significant improvement over NBE under correct
specification,
significantly outperforms NBE under contamination,
and achieves by far the best performance under dependence misspecification.
These results suggest that the adaptive Wasserstein refinement can be useful in heavy-tailed settings, especially when the reference model is misspecified. In this experiment, AW--NBE improves over NBE in all three Double Pareto scenarios, with the clearest gain observed under dependence misspecification.

\section{Overall: discussion and key takeaways}
\label{sec:conclusion_benchmark}
From this empirical study, several conclusions emerge.
First, no estimator dominates uniformly across all settings. The MLE achieves
the lowest RMSE under correct specification, as expected from likelihood-based
estimation in well-specified regular models. However, its performance deteriorates
substantially under contamination, overdispersion and dependence misspecification.
This fragility reflects the fact that the likelihood criterion is tied to the
assumed parametric form; when this form is misspecified (as is often the case with real data), the resulting optimum
may no longer correspond to the target component of the data-generating
mechanism.

Second, EOT provides strong robustness in several misspecified scenarios, but
this robustness comes at the cost of efficiency under correct specification.
The Sinkhorn divergence evaluates discrepancy at the distributional level,
rather than through an explicit likelihood. This makes it less dependent on
the exact parametric density assumed by the model. At the same time, its
performance depends on regularization, scaling and preprocessing choices,
especially in heavy-tailed settings.

Third, NBE occupies an intermediate position. It provides stable and fast point
estimation once trained, but it can degrade under distributional shift. This is
expected, since the estimator is trained under a reference simulator and its
squared-error training objective reflects the parameter-data relationship
induced by that simulator. When the observed data depart from the training
distribution, the learned map may no longer extrapolate reliably.
\\[1ex]
To compare the estimators across all experiments, we now introduce a set of aggregate criteria that summarize performance across the nine scenarios. These criteria combine per-scenario RMSE ranks, the Borda count \citep{borda1784memoire}, the number of first- and last-place rankings, and relative performance gaps. 
The relative performance gap follows the performance-ratio normalization used in performance profiles
\citep{dolan2002benchmarking}.
For scenario $s$ and method $k$, it is defined as
\[
\mathrm{RelGap}_{s,k}
=
\frac{\mathrm{RMSE}_{s,k}}
{\min_{\ell}\mathrm{RMSE}_{s,\ell}}
-1.
\]
This quantity measures the proportional excess RMSE of method $k$ relative to the best-performing method in the same scenario. Table~\ref{tab:Overall:_ranking} summarizes the resulting comparisons.
\begin{table}[H]
\centering
\caption{\sf \small Overall: ranking of point estimation methods across all models 
and scenarios. Ranks 1 (best) to 4 (worst) are based on 
RMSE$(\widehat{\boldsymbol{\theta}})$. RMSE values are given in 
parentheses. \textbf{Bold} indicates the best method per row.
Misspecification 1: inflated variance / Poisson--Gamma ($r=5$) / 
heavy-tail contamination. Misspecification 2: Student-$t$ ($\nu=3$) / 
Poisson--Gamma ($r=20$) / Student-$t$ copula.
The relative performance gap is defined as 
$\mathrm{RelGap}_{s,k} = \mathrm{RMSE}_{s,k}/\min_{\ell}\,
\mathrm{RMSE}_{s,\ell} - 1$; mean and worst-case are computed 
over all nine scenarios.}
\label{tab:Overall:_ranking}
\resizebox{\textwidth}{!}{%
\begin{tabular}{ll|cccc}
\toprule
\textbf{Scenario} & \textbf{Model}
  & \textbf{MLE} & \textbf{NBE} & \textbf{EOT} & \textbf{AW--NBE} \\
\midrule
\multirow{3}{*}{Well-specified}
  & Gaussian AR(1)
    & $\mathbf{1}\ (0.327)$ & $2\ (0.371)$ & $4\ (0.521)$ & $3\ (0.410)$ \\
  & Poisson common shock
    & $\mathbf{1}\ (0.567)$ & $3\ (0.655)$ & $4\ (0.738)$ & $2\ (0.646)$ \\
  & Double Pareto
    & $\mathbf{1}\ (0.175)$ & $3\ (0.574)$ & $4\ (1.174)$ & $2\ (0.559)$ \\
\midrule
\multirow{3}{*}{Misspecification 1}
  & Gaussian AR(1)
    & $4\ (1.804)$ & $3\ (1.364)$ & $\mathbf{1}\ (0.643)$ & $2\ (0.991)$ \\
  & Poisson common shock
    & $3\ (1.943)$ & $4\ (2.165)$ & $\mathbf{1}\ (1.403)$ & $2\ (1.799)$ \\
  & Double Pareto
    & $4\ (0.788)$ & $3\ (0.717)$ & $\mathbf{1}\ (0.592)$ & $2\ (0.611)$ \\
\midrule
\multirow{3}{*}{Misspecification 2}
  & Gaussian AR(1)
    & $4\ (2.403)$ & $3\ (1.905)$ & $\mathbf{1}\ (1.033)$ & $2\ (1.526)$ \\
  & Poisson common shock
    & $4\ (1.850)$ & $3\ (1.692)$ & $\mathbf{1}\ (1.354)$ & $2\ (1.659)$ \\
  & Double Pareto
    & $4\ (0.645)$ & $2\ (0.588)$ & $3\ (0.594)$ & $\mathbf{1}\ (0.306)$ \\
\midrule\midrule
\multicolumn{2}{l|}{\textbf{Borda count (\cite{borda1784memoire})}}
    & $2.89$ & $2.89$ & $2.22$ & $\mathbf{2.00}$ \\
\multicolumn{2}{l|}{\textbf{Rank 1 count}}
    & $3$ & $0$ & $5$ & $1$ \\
\multicolumn{2}{l|}{\textbf{Rank 4 count}}
    & $5$ & $1$ & $3$ & $0$ \\
\multicolumn{2}{l|}{\textbf{Mean rel.\ performance gap} $\downarrow$}
    & $0.59$ & $0.72$ & $0.84$ & $\mathbf{0.46}$ \\
\multicolumn{2}{l|}{\textbf{Worst-case rel.\ performance gap} $\downarrow$}
    & $\mathbf{1.81}$ & $2.28$ & $5.71$ & $2.19$ \\
\bottomrule
\end{tabular}%
}
\end{table} 
Table~\ref{tab:Overall:_ranking} confirms that no single method dominates across all settings. The MLE and EOT represent two extremes of the  efficiency--robustness trade-off: the MLE excels under correct specification 
but is the most fragile under misspecification, whereas EOT shows the opposite pattern. NBE, despite its computational advantages once trained, offers no clear improvement over MLE in terms of robustness, as illustrated by its severe degradation under Poisson--Gamma overdispersion. 

AW--NBE occupies a different position. It is the only method never ranked last, achieves the lowest Borda count ($2.00$), and attains the smallest mean relative performance gap ($0.46$). Rather than optimizing performance in a single setting, it provides the most stable compromise across well-specified and misspecified experiments. These results suggest that robustness in likelihood-free point
estimation depends not only on the neural architecture itself, but also on the criterion used at inference time to compare observed and simulated data. From this perspective, adaptive combinations of neural estimation and optimal transport refinement offer a promising balance between statistical efficiency and stability under misspecification.

This conclusion should nevertheless be understood as criterion-dependent:
AW--NBE is not claimed to be uniformly optimal, but rather to provide the most
consistent overall performance across the models and scenarios considered here.
\\[1ex]
Nevertheless, several aspects of this benchmark invite further investigation. 
For computational feasibility, the experiments are restricted to trivariate models with a fixed sample size of $m = 50$. Even under these constraints, the total runtime reached approximately 7.5 days on a standard laptop, as reported in Section~\ref{sec:framework}.
It therefore remains an open question how the relative rankings between estimators evolve in higher dimensions or for larger datasets. The hyperparameters of the neural and transport-based estimators are also held fixed across all scenarios, and adaptive tuning could improve their performance under misspecification. These choices reflect a deliberate focus on reproducibility and interpretability rather than exhaustive coverage.
\\[1ex]
%%
% Overall, these results suggest that robustness in likelihood-free point estimation depends not only on the estimator's training objective but also on the discrepancy criterion applied at inference time. Among the methods considered, AW--NBE achieves the most consistent compromise between statistical efficiency under correct specification and stability under misspecification, supporting adaptive combinations of neural and optimal-transport-based estimation as a promising direction for robust simulation-based inference.
%%
Taken together, the experiments indicate that robust likelihood-free point estimation depends not only on the estimator architecture, but also on the discrepancy criterion used at inference time. Within the settings considered here, adaptive combinations of neural estimation and optimal transport provide a promising compromise between efficiency and robustness.
%%%%
%%%%
\paragraph{Acknowledgement.} 
The work of S. Aka was supported by ANRT (CIFRE PhD program) in collaboration with {\it Square Management}.
\\P. Naveau acknowledges financial support from the ANR through the SICIM and SHARE PEPR Maths-Vives projects (France 2030, ANR-24-EXMA-0008), the EXSTA grant (ANR-23-CE40-0009-01), PORC-EPIC, the PEPR TRACCS program (PC4 EXTENDING, ANR-22-EXTR-0005), and the PEPR IRIMONT project (France 2030, ANR-22-EXIR-0003). He also benefited from the GeoLearning research chair.
\bibliographystyle{apalike}
\bibliography{Ref}
\end{document}